# Uncovering Long Memory in High Frequency UK Futures

**JOHN COTTER**

**University College Dublin**

**Address for Correspondence:**

Dr. John Cotter,

Centre for Financial Markets,

University College Dublin,

Blackrock,

Co. Dublin,

Ireland.

E-mail. john.cotter@ucd.ie

Ph. +353 1 716 8900

Fax. +353 1 283 5482

The author would like to thank two anonymous referees, the editor and participants at Forecasting Financial Markets 2002, and the International Symposium on Forecasting 2002 for their helpful comments. University College Dublin Faculty research funding is gratefully acknowledged.

# Uncovering Long Memory in High Frequency UK Futures

**Abstract:**

Accurate volatility modelling is paramount for optimal risk management practices. One stylized feature of financial volatility that impacts the modelling process is long memory explored in this paper for alternative risk measures, observed absolute and squared returns for high frequency intraday UK futures. Volatility series for three different asset types, using stock index, interest rate and bond futures are analysed. Long memory is strongest for the bond contract. Long memory is always strongest for the absolute returns series and at a power transformation of k < 1. The long memory findings generally incorporate intraday periodicity. The APARCH model incorporating seven related GARCH processes generally models the futures series adequately documenting ARCH, GARCH and leverage effects.

**Uncovering Long Memory in High Frequency UK Futures**

# 1 Introduction

Volatility is a latent variable fundamental to asset pricing, asset allocation and risk management. For instance, the quality of risk management practices is critically determined by the modelling of financial volatility and its associated attributes must be measured and modelled properly. An extensive literature has characterized the systematic properties of volatility at the daily and lower frequencies for equity and fixed-income assets. Here, volatility is both time-varying and predictable and this latter feature importantly gives rise to long memory where persistence occurs for large lags. This property is important for risk management as it affects the monitoring and management of risk associated with market trading.

This paper investigates long memory in alternative risk measures, observed absolute and squared returns for high frequency futures data.[1] The paper determines which risk measure exhibits long memory at it strongest in terms of length and magnitude of persistence. Whilst the long memory property is cited in the literature for exchange rate and equity series this paper extends the analysis to less volatile assets, using interest rate and bond futures series. Furthermore, the paper investigates long memory at relatively high frequency intraday intervals of interest to the everyday operations of a trading desk. 5-minute intervals are chosen to minimise non-synchronous trading effects.

---

[1] Granger (1998) notes that long memory is usually discussed in the context of squared returns series, but that absolute returns series have more interesting statistical properties thus motivating the investigation in this study.

The long memory property occurs where volatility persistence remains at large lags as in an absolute returns series, $|R_t|^k$, or a squared returns series, $[R_t]^k$, where $k > 0$. Long memory is documented for daily equity prices series where volatility persistence decays relatively slowly for a long period after an initial rapid decline (Ding et al, 1993; Ding and Granger, 1996). In addition, Ding et al (1993) indicate that this non-linear dependence is strongest at the power transformation of $k = 1$ suggesting that volatility modelling incorporating this property should focus at the level of returns rather than squared returns. For intraday currency realisations, previous evidence shows that the slow decay of the autocorrelation structure involves a u-shaped cyclical pattern describing intraday volatility behaviour (Andersen and Bollerslev, 1997a, 1997b).

This paper examines whether these characteristics of long memory are evident for high frequency bond, interest rate and equity futures and what variations occur (if any) according to asset type. The paper determines which power transformation exhibits long memory at its strongest, and whether intraday cycles exist for the asset types analysed.

In terms of model building, parametric ARCH models have emerged as the archetype for modelling time-varying **and predictable** volatility. There are a large number of possible specifications available incorporating many stylized features of financial returns. One such model proposed by Ding et al (1993), the Asymmetric Power ARCH (APARCH), has considerable success in modelling time-varying and predictability features of daily returns is applied here to the high frequency series. Furthermore the simulated autocorrelation function mirrors the long memory features of their daily returns series. This model encompasses seven different GARCH related specifications incorporated into a single model nesting ARCH, GARCH, and leverage effects coupled with different power transformations of the volatility process (see Shephard, 1996; for a survey of time-varying volatility models and their applications). The model is applied here to determine what stylized features of volatility are associated with the high frequency futures and whether this family of GARCH models adequately models intraday volatility effects.

The paper proceeds as follows. In section 2, long memory is discussed. The section is completed by a presentation and discussion of a single representation of seven GARCH related processes fitted to the intraday futures series. Details of the futures contracts chosen and data capture follow in section 3. Section 4 presents the empirical findings. It begins by a thorough analysis of the futures long memory characteristics. In addition, the stylized features of high frequency futures series are documented from fitting the APARCH process. Finally, a summary of the paper and some conclusions are given in section 5.

**2 Theory of Long Memory and Financial Volatility:**

**2.1 Long Memory:**

Long memory properties may be investigated by focusing on the absolute returns series, $|R_t|^k$, or the squared returns series, $[R_t^2]^k$, and on their power transformations, where $k > 0$.[2] Models with a long memory property have dependency between observations of a variable for a large number of lags so that Cov[$R_{t+h}$, $R_{t-j}$, $j \geq 0$] does not tend to zero as h gets large. In contrast, if the dependency between observations of a variable disappears for a small number of lags, h, such as for a stationary ARMA process then the model is described as having a short memory property and Cov[$R_{t+h}$, $R_{t-j}$, $j \geq 0$] $\to$ 0. Formally, long memory is defined for a weakly stationary process if its autocorrelation function ρ(·) has a hyperbolic decay structure:

$$\rho(j) \sim Cj^{2d-1} \text{ as } j \to \infty, C \neq 0, 0 < d < \frac{1}{2}, \qquad (1)$$

---

[2] For an excellent treatment of long memory processes see Beran (1994). An alternative approach to this paper that does not examine the characteristics of long memory for volatility series calculates the degree of fractional integration, d, and this is generally found to be close to 0.4 (Taylor, 2000).

Baillie (1996) shows that long memory processes have the attribute of having very strong autocorrelation persistence before differencing, and thereby being non-stationary, whereas the first differenced series does not demonstrate persistence in themselves and is stationary. However, the long memory property of these price series is not evident from just first differencing alone, but has resulted from analysis of risk measures. Long memory has been documented across a large sphere of the finance literature from macroeconomic series such as GNP (Diebold and Rudebusch, 1989) to exchange rate series (Dacorogna et al, 1993; Baillie et al, 1996; Andersen and Bollerslev, 1997a, 1997b). Of closer relevance to this study long memory is documented for equity index series, albeit at daily intervals (Ding et al, 1993, Ding and Granger, 1996).

The explanations for long memory are varied. One economic rational results from the aggregation of a cross section of time series with different persistence levels (Andersen and Bollerslev, 1997a; Lobato and Savin, 1998). Alternatively, regime switching may induce long memory into the autocorrelation function through the impact of different news arrivals (Breidt et al, 1998). The corresponding shape of the autocorrelation function may be hyperbolic, beginning with a high degree of persistence that reduces rapidly over a few lags, but that slows down considerably for subsequent lags to such an extent that the length of decay remains strong for a very large number of time periods. Also, with a slight variation, it may follow a slowly declining shape incorporating cycles that correspond to intraday volatility patterns associated with different trading hours (Andersen and Bollerslev, 1997a, 1997b).

Whilst second order dependence is a characteristic of financial returns, usually modeled by a stationary GARCH process, these specifications have been questioned as to their ability to model the long memory property adequately in contrast to their Fractionally Integrated GARCH counterparts (Baillie, 1996). For instance, while stationary GARCH models show the long memory property of financial returns volatility series occurs by having $[R_t^2]$ and $|R_t|$ with strong persistence, they assume that the autocorrelation function follows a certain pattern not corresponding to a long memory process. In particular, the correlation between $[R_t^2]$ and $|R_t|$ from stationary GARCH models and their power transformations remain strong for a large number of lags, with the rate of decline following a constant pattern (Ding et al, 1993), or an exponential shape (Ding and Granger, 1996). In contrast, a number of returns series, both $[R_t^2]$ and $|R_t|$, in fact have been found to decay in a hyperbolic manner, namely, they decline rapidly initially, and this is followed by a very slow decline (Ding and Granger, 1996). However, Ding et al (1993) find that the APARCH model nesting seven GARCH specifications adequately models long memory for equity series observed at daily intervals and this analysis is extended for relatively high frequency intraday series.[3]

## 2.2 APARCH Model:

Ding et al (1993) propose a generalised version of seven GARCH related processes with a link based on their parameter values, named an APARCH model. This model nests the following specifications: ARCH (Engle, 1982); Non-linear ARCH - NARCH (Higgins and Bera, 1992); Log-ARCH (Geweke, 1986/Pantula, 1986); GARCH with variance (Bollerslev, 1986); GARCH with standard deviation (Taylor, 1986/Schwert, 1990), Threshold ARCH - TARCH (Zakoian, 1991); and GJR (Glosten et al, 1993). In fitting the APARCH model to time series, it offers the flexibility of dealing with different power transformations incorporating for instance, the variance and standard deviation that are associated with the identification of long memory for financial data. Furthermore by incorporating Schwert's (1990) model it allows for absolute realisations, again used to model long memory features of financial returns. Given the success of applying the APARCH specification in modelling long memory at daily intervals (Ding et al, 1993), it is interesting to extend this analysis to intervals of interest to the everyday operations of a trading desk.

The APARCH model, will in the first instance, be applied to determine which features of the seven processes describe the volatility characteristics of the futures data. Secondly it will examine the features of the standardised residuals from fitting the APARCH model. As well as describing the traditional time dependent volatility feature, the model specifically incorporates the leverage effects (see Black, 1976), by letting the autoregressive term of the conditional volatility process be represented as asymmetric absolute residuals. The most appropriate version from these seven processes is determined through statistical analysis of the fitted APARCH model. The volatility expression is given as:

$$r_t = \mu + \sigma_t \varepsilon_t$$
$$\varepsilon_t = r_t - \mu$$
$$\sigma_t^\delta = \alpha_0 + \sum_{i=1}^{p} \alpha_i (|\varepsilon_{t-i}| + \gamma_i \varepsilon_{t-i})^\delta + \sum_{j=1}^{q} \beta_j \sigma_{t-j}^\delta \quad (2)$$
$$\text{for } \alpha_0, \alpha_i, \beta_j \geq 0,\ \alpha_i + \beta_j \leq 1,\ -1 \leq \gamma_i \leq +1$$

---

[3] Alternative modelling approaches using for example, Fractionally Integrated GARCH and Long Memory Stochastic Volatility processes could also be followed but are not examined in this study.

The residuals, $\varepsilon_t$, were initially assumed to be from a conditionally normal process as in Engle's (1982) representation of the stochastic volatility process, but can easily take on other conditional assumptions. Given the commonly found fat-tailed characteristics in financial returns, the conditional mean process can be adjusted for this attribute with the assumption of a student-t distribution as in (Baillie and DeGennaro, 1990) or the generalised exponential distribution as in (Nelson, 1989).

**3 Data Considerations**

The empirical analysis is based on transaction price data obtained from *Liffedata* for futures contracts trades on the LIFFE exchange.[4] The FTSE100 stock index contract, the UK Long Gilt bond contract, and the Three-month Sterling interest rate contract are the asset type proxies chosen for analysis between 1998-1999.

---

[4] Information is available on the time of trading upto the nearest second, the originator of the trade whether from the trading floor or electronically, the price and volume traded of the contract and its expiry date, and the transaction code (bid, ask, trade, spread and volatility). Furthermore, this exchange has made a clear distinction, between contracts that are either linked to an underlying asset or developed formally on the basis of links to the recently developed European currency, the Euro, and those that remain linked to factors outside the currency area. Representative samples of the latter asset type based on the consideration of being the most actively traded, and thus providing more accurate information, using market volume are included.

Data is available on contracts for four specific delivery months per year, March, June, September and December with prices chosen from the most actively traded delivery months using a volume crossover procedure. The empirical analysis is completed on the futures contracts for a sampling frequency of five-minute prices based on microstructure arguments. Primarily, the bias induced by non-synchronous trading as a result of interpolation of tick data at this time interval is minimised (Andersen et al, 1999). This non-synchronous trading issue results in large and negative autocorrelation in the newly formed returns' series (Lo and MacKinlay, 1990). As an example, consider dependency in tick data caused by bid-ask bounces which may be as a result of the sequential execution of limit orders on the books of a specialist as the market moves through those limit prices. Turning to less high frequency intervals, for example 5-minutes, minimises this effect. For each contract, log closing prices (or log closing prices to the nearest trade available) for each interval are first differenced to obtain each period's return.

A full trading day, and consequently the full set of returns, depends on hours of trading and holiday effects. Each futures contract is cleaned for these effects, as they would impact on respective contracts' time series characteristics. In particular, the FTSE100 future's trading day is between 08.35 and 18.00, the UK Long Gilt between 08.00 and 18.00, and the Sterling contract between 08.05 and 17.55.[5] Prior to cleaning, the futures (FTSE100 – 113, UK Long Gilt – 120 and Sterling – 118) involved different numbers of daily trading intervals. Turning to the specific details of the data capture, trading does not take place during weekends. In addition all contracts had holidays removed involving 9 per year.

---

[5] The last trade on the Sterling contract is actually 17.57 but the cut-off is imposed due to the lack of a complete five-minute interval.

These consisted of New Year's (two days), Easter (two days), May Day, Spring holiday (1 day), Summer holiday (1 day), and Christmas (two days). This results in 251 trading days per year[6]. Finally, any five-minute interval not including a trade is excluded, and this had the greatest impact on the Sterling contract.[7] For instance, relatively scant trading volume occurs for the Sterling futures between 12 and 2pm.

**4 Empirical Findings**

**4.1 Long Memory Properties of UK Futures:**

A preliminary examination of the futures data indicates a number of well-known characteristics. These include non-symmetric leptokurtotic returns. Furthermore, excess kurtosis becomes more pronounced moving from the returns series to the volatility estimates.[8] The autocorrelation function (ACF) is used to determine dependency in the UK futures series following amongst others Ding et al (1993) and Granger and Ding (1996) over a large number of lags.[9] The patterns of the autocorrelation values have similar features for the different assets analysed. Generally we see that the autocorrelation values of returns are negligible compared to absolute and squared returns series suggesting that dependence is not evident in returns themselves but rather is prevalent in the volatility series. In particular this is true for the UK Long Gilt contract that has relatively few significant autocorrelation values for the returns series, but very strong persistence evident for the volatility series at very large lag numbers. This second stylised fact documents the long memory property of the volatility series having an autocorrelation structure that decays slowly, with significant positive values over a large number of lags.

---

[6] Specifically, in 1998 this involved January 1, April 10, April 13, May 4, May 25, August 31, December 25, December 28, December 31; and in 1999 it involved January 1, April 2, April 5, May 3, May 31, August 30, December 27, December 28, December 31.
[7] In contrast, many intervals contained a multitude of trades although only a single return is computed for each interval. However, it is important to note that the long memory properties may be affected by thin trading, especially the Sterling contract. The cleaned series total 55011 (FTSE100), 57227 (UK Long Gilt), and 27406 (Sterling) returns respectively.
[8] The preliminary findings are not presented to aid conciseness and are available on request.
[9] Although there is possible miss-specifications with the approach examining only a few lags (Lobato and Savin, 1998), the extensiveness of the analysis are able to clearly identify the long memory property. Furthermore, alternative approaches for identifying long memory such as the LM test and Wald test are invalidated if a finite fourth moment does not exist (Lobato and Savin, 1998) as indicated for financial returns by Loretan and Phillips (1993).

As previous studies (see for example, Taylor, 1986) suggest that the long memory of low frequency returns series tend to be stronger for the absolute returns series at the expense of the squared series, it is worth analysing this issue for high frequency intervals. In table 1, we see the number of significant autocorrelation values for each volatility series incorporating separate power transformations. The findings are clear with few exceptions, notably, the assets analysed here confirm the previous findings that the persistence property is strongest for the absolute returns series.

INSERT TABLE 1 HERE

Taking the bond contract, the UK Long Gilt future as an example to illustrate variations in dependency, we see that squared and absolute values are reasonably similar in magnitude (0.2455 versus 0.2399 for the series $k = 1$, and at lag 1), but that the latter values generally dominate the former. The extent of the long memory property for this futures contract regardless of the volatility measure is clearly evident with an autocorrelation structure that decays very slowly with all 5722 lags statistically significant for $k = 1$. Similar findings are made for the other contracts. Notable exceptions include the Sterling contract with the power transformation k = 0.25 at almost all lags, and the FTSE100 contract for the equivalent series at lags 1 and 2. The values in table 1 also demonstrate the relative strength of long memory according to asset type. From all the results, it is clear that the long memory property is strongest for the UK Long Gilt contract, with the characteristic weakest for the Sterling contract. The ambiguity regarding the long memory property, and also as we will see later its intraday patterns for the Sterling contract may be as a result of its degree of thin trading especially around lunchtime vis-à-vis the other contracts.

A further issue that invites examination is which power transformation, k, demonstrates the long memory property at its strongest. To answer this, we again examine the values in table 1. Interestingly, this property is never at its strongest for the original volatility series, namely where k = 1. For example, taking the squared returns series of the FTSE100 as a case in point, we see that persistence is strongest for the power transformation of k = 0.25 for nearly all lags (4598) significant. Concentrating on the absolute returns series that dominate the autocorrelation values in terms of magnitude, we see that long memory is strongest for the stock index contract at k = 0.5, the bond contract at k = 0.5 (with larger autocorrelation values for respective lags), and the Sterling contract at k = 0.75 according to the aggregate levels of slow autocorrelation decay, as well as the magnitude of the individual lag values cited. In contrast, Ding et al (1993) find that autocorrelation is strongest for k = 1 for stock indexes using daily observations. The associated correlation values for these series showing the long memory property at its strongest shows the full extent of persistence in the assets analysed with the UK Long Gilt having all lags positive and significant, the FTSE100 being only slightly less resolute for the aggregate numbers of significant lags and in terms of magnitude, and finally, the Sterling contract's long memory results diverging from the other two assets. This latter asset's findings are surprising given the general support of long memory offered for financial time series in general, but it does point out that asset type classification may be an important determinant in citing persistence and its associated degree of strength.

Having first established the long memory property of volatility measures, it is worth exploring the shape of the autocorrelation function to determine their patterns. Previous evidence suggests that the autocorrelation function for asset volatility series follows a hyperbolic decay structure that decreases relatively fast initially, and then starts decreasing very slowly (Ding et al, 1993). In figure 1 each respective contract's autocorrelation values are plotted over a trading week where the long memory property is strongest, namely for the absolute returns series representing 5 complete trading days. Certainly this appears to be true for the UK Long Gilt Contract with for example, the first 20 lags of untransformed absolute returns series falling consecutively from 0.2455 to 0.0454 and then decays slowly. Similarly, this is also true for the Sterling contract, the one that shows the weakest signs of long memory persistence. However, there are a number of discrepancies here that suggest further examination. For instance, looking at the FTSE100, we see a general decay in autocorrelation values at greater lags, but this does not follow the hyperbolic decay structure. Here, there appears to be a cyclical pattern within the declining persistence, and this would agree with evidence presented for stock index and exchange rate series (Andersen and Bollerslev 1997a; 1997b).

INSERT FIGURE 1 HERE

Daily periodicity of persistence is clearly evident for the stock index and bond futures with 5 repeated cycles presented. In contrast, no clear pattern emerges for the interest rate contract, with the exception that persistence is reasonably small, especially after lag 100 where the series resembles realisations of white noise. Specifically, for the FTSE100 contract, volatility persistence follows the u-shaped pattern corresponding to high levels at opening and closing times that surround lower levels during the rest of the day. Support for this pattern in the volatility measures are caused by strategic interaction of traders at opening and closing times is offered for other stock indexes (Brock and Kleidon, 1992). The pattern for the periodicity of the persistence of the UK Long Gilt differs slightly, with the existence of an overall u-shape that incorporates two smaller u-shaped patterns involving half-day cycles. Also, there is a slight day-of-the week effect for these latter two assets with an overall consecutive reduction in the first four cycles followed by a small increase in the peak on the fifth cycle.[10]

**4.2 APARCH Model Findings:**

Turning to the GARCH modelling, each returns series is fitted with a multitude of APARCH specifications using the Berndt-Hall-Hall-Hausman (1974) algorithm in S-PLUS. Variations under the auspices of modelling the conditional mean with an ARMA (and separately with an AR and MA) process, coupled with various numbers of lags (P, Q) in the conditional variance expression, are examined. In addition, to assuming that the conditional returns series can be modelled with gaussian errors, the commonly noted fat-tailed characteristic of financial returns is accounted for, by modelling the error terms with student-t and generalised exponential distributions. Findings for the optimal models based on the AIC and BIC criteria are presented in table 2 offering support for an APARCH specification. The APARCH model itself is well specified with all parameters significant at asymptotic significance levels. Generally, the conditional volatility models are similar in their attributes. For each series, the APARCH specification relying on a conditional student-t distribution dominates the other models. The parsimonious specification with p and q equalling one is also optimal for each series. Inclusion of AR and MA terms in the conditional mean equation is valid for the Long Gilt and Sterling contracts, whereas neither term is appropriate for the FTSE100 futures. Also, significant leverage effects, $\gamma$, are accepted for each contract, demonstrating that information has an asymmetric effect on volatility with bad news having a greater impact than good news.

---

[10] The pattern for the FTSE100 is very similar to the S&P 500 (Andersen and Bollerslev, 1997b). In contrast, weekend effects would also be evident in the u-shaped pattern for assets that trade continuously over the full trading week (Dacorogna et al, 1993).

INSERT TABLE 2 HERE

Whilst all contracts exhibit similar time series characteristics, with for example, all confirming the covariance stationary property, by having $\alpha + \beta < 1$, there is some divergence in the volatility estimates with the Sterling contract exhibiting much smaller but significant persistence with its $\beta$ estimate. However in general, the volatility parameter estimates for 5-minute intervals correspond to those from daily observations. Whilst changing persistence parameters have been noted for estimation at different frequencies daily and 5-minute estimates are similar (McMillan and Speight, 2002). An implication of these parameter estimates investigates the volatility clustering for each futures examining the half-life of the impact of a shock to the volatility measure, $\sigma_t$, calculated from $-\log 2(\log \alpha_i + \beta_i)^{-1}$. This measure gives the number of time periods it takes for half the expected reversion back to $\sigma_t$ after a shock occurs. From table 2, we see that on average the persistence in the volatility measures for the UK Long Gilt and FTSE100 contracts are the same at approximately 114 minutes (22.75 5-minute intervals). These differ substantially from the smaller persistence of the Sterling contract's value of 13 minutes (2.65 5-minute intervals).

As volatility persistence is found for all futures series, a concise method of distinctions is made as to the most appropriate version of GARCH related specification by examining the coefficients of the volatility model. Here we see, in addition to ARCH effects, GARCH and leverage effects also occur. The significant leverage effects illustrate that information has an asymmetric effect on volatility. Thus Engle's (1982) ARCH, Higgins and Bera (1992) Nonlinear ARCH, and the Log-ARCH process of Geweke (1986) and Pantula (1986) would not optimally fit each series volatility generating process due to the lack of GARCH and leverage characteristics. In addition, the lack of moving average terms, $\beta_j$, in the volatility expression outweighs the advantages of the inclusion of a leverage term in Zakoian's (1991) TARCH model. Similarly, whilst GARCH effects are documented for each series, the existence of statistically significant leverage effects are included in neither Bollerslev's (1986) or Taylor (1986) and Schwert's (1990) generalised processes. Moving to the final model that is incorporated by an APARCH process, there does appear to be support initially for GJR's (1993) model with the inclusion of both ARCH and GARCH effects, coupled with asymmetric leverage effects. However, as the t-statistics of $\delta$, are all significantly different from two, the power characteristic of GJR's model is not accepted. In fact, whilst the APARCH process is itself well specified with all parameters significant, there is no clear conclusion with regards to the suitability of the specific nested models $\delta$ values for any of the futures contracts analysed.

As stated, the second objective from fitting the APARCH model is to determine the features of the standardised residuals, the residuals divided by the APARCH estimates of conditional volatility. Autocorrelation findings are presented in table 3 and we see that the rescaling does not have a consistent effect on the original series. For instance, persistence is strongest both in terms of magnitude and staying power for the absolute standardised residuals series. In addition, long memory is never strongest for the standardised residuals volatility series at $k = 1$ regardless of whether it's the absolute or squared values. However, in comparing the absolute series with the most persistence, we see a slight change. Whilst the findings for the stock index contract remain unchanged ($k = 0.5$), the magnitude of autocorrelation values for the UK Long Gilt series, $k = 0.25$, now outweighs its original counterpart. Also, there is an impact for the Sterling contract, with $k = 0.25$ now offering the strongest levels of persistence, and a very large first spike (0.4258).

INSERT TABLE 3 HERE

Focusing more closely on the findings in table 3, we do see some minor differences in the autocorrelation structure of the standardised residuals volatility series. For the squared standardised residuals, the persistence has a quicker decay after fitting the conditional volatility model. For example for the series $k = 0.25$, the full impact is complete after 2224 lags, whereas it remains in the absolute returns series for the full lifetime of analysis. Also, for the squared standardised residuals, there are slightly weaker autocorrelation values for the Sterling contract at d = 1. In contrast, differences in the absolute standardised residuals series are mainly in having stronger dependence for the contract analysed. For example with the FTSE100, there is a similar structure to the decay of the autocorrelation values for the standardised residuals series vis-à-vis the original series, but with the long memory property being slightly more pronounced for the former series. The strongest signs of this characteristic occur for k = 0.25 after obtaining the standardised residuals from fitting the APARCH specification. Generally the rescaling does not remove long memory indicating an inability of the APARCH model to capture this property for high frequency interval.

## 5 Summary and Conclusions

Accurate volatility measures are paramount for optimal risk management practices. The intraday features in the risk measures have important implications for modelling the volatility of high frequency realisations. However intuitively, financial markets and the behaviour of volatility patterns would have a number of sources of time dependence, for example cyclical occurrences. This paper concentrates on identifying and accounting for the long memory property. First, long memory is investigated through the autocorrelation function for a large number of lags of two volatility series, absolute returns, and squared returns. Then, seven related GARCH processes through an APARCH specification, are fitted to the data to determine what stylized features are inherent in the high frequency realisation. In particular, previous successful evidence in modelling long memory using the APARCH process at daily intervals is the backdrop for the analysis in this study. This paper turns its attention to intraday volatility series for three different asset types, using stock index, and the less risky, interest rate and bond futures. The FTSE100, UK Long Gilt and Sterling contracts are the most actively traded representatives of each asset type from the LIFFE exchange. Observations at 5-minute intervals are chosen to minimise non-synchronous trading effects. Each contract is cleaned to remove non-trading and holiday effects.

The results presented for the volatility series are novel and interesting. First, as well as volatility magnitudes varying by asset type, long memory properties do also. In particular, dependency in the volatility structure at a large number of lags is strongest for the bond contract, followed by the stock index contract. Second, long memory is strongest for the absolute returns series for all contracts. This feature occurs in the length of the memory and the magnitude of the dependence. Third, for all assets, the long memory property is strongest at a power transformation of $k < 1$. These two findings suggest that the use of absolute returns as a volatility measure and at different transformations offer attractive alternatives for describing the long memory property. Fourth, intraday periodicity is strongly supported for two of the assets analysed, with again the interest rate futures displaying weaker evidence.

Turning to the findings from fitting the parametric APARCH model, further important conclusions are made. Whilst the APARCH model itself fits the data adequately in terms of describing the general volatility features of the data, none of the seven separate GARCH models are fully well specified. This implies that modelling volatility with a generalised process incorporating a number of stylized features may dominate modelling with individual standard GARCH related specifications. Furthermore, the APARCH process is unable to remove the long memory features by rescaling the original futures series. Long memory remains, and follows a slightly different pattern prior to rescaling. Future parametric work on high frequency realisations should incorporate alternative approaches such as the discrete Fractionally Integrated GARCH (FIGARCH) related models (Baillie et al, 1996) or the Long Memory Stochastic Volatility (LMSV) processes (Breidt et al, 1998) to assess their ability to capture long memory with cyclical intraday patterns.

Ding, Z., C. Granger, C. W. J., and R. F. Engle (1993) A Long Memory Property of Stock Returns, Journal of Empirical Finance, 1, 83-106.

Ding, Z., and C. W. J. Granger (1996) Modelling Volatility Persistence of Speculative Returns, Journal of Econometrics, 73, 185-215.

Engle, R. F. (1982) Autoregressive Conditional Heteroskedasticity with Estimates of the Variance of UK Inflation, Econometrica, 50, 987-1008.

Geweke, J. (1986) Modelling the Persistence of Conditional Variances: A Comment, Econometric Reviews, 5, 57-61.

Glosten, L., R. Jaganathan and D. Runkle (1993) On the Relations between the Expected Value and the Volatility of Nominal Excess Return on Stocks, Journal of Finance, 48, 1779-1801.

Granger, C. W. J. (1998) Comment on 'Real and Spurious Long-Memory Properties of Stock-Market Data', Journal of Business and Economic Statistics, 16, 268-269.

Higgins, M. and A. Bera (1992) A Class of Nonlinear ARCH Models, International Economic Review, 33, 37-158.

Lo, A. and A. C. MacKinlay (1990) An Econometric Analysis of Nonsynchronous Trading, Journal of Econometrics, 45, 181-212.

Lobato, I. N. and N. E. Savin (1998) Real and Spurious Long-Memory Properties of Stock-Market Data, Journal of Business and Economic Statistics, 16, 261-268.

Loretan, M., and P. C. B. Phillips (1993) Testing the Covariance Stationarity of Heavy-Tailed Time Series: An Overview of the Theory with Applications to Several Financial Datasets, Journal of Empirical Finance, 1, 211-248.
23

McMillan, D. G. and A. E. H. Speight (2002) Temporal Aggregation, Volatility Components and Volume in High Frequency UK Bond Futures, The European Journal of Finance, 8, 70-92.

Nelson, D. B. (1989) Modelling Stock Market Volatility, Proceedings of the American Statistical Association, Business and Economics Statistics Section, pp. 93-98.

Pantula, S. G. (1986) Modelling the Persistence of Conditional Variances: A Comment, Econometric Reviews, 5, 71-73.

Schwert, G. W. (1990) Stock Market Volatility and the Crash of '87, Review of Financial Studies, 3, 77-102.

Shephard, N. (1996) Statistical Aspects of ARCH and Stochastic Volatility, in D. R. Cox, D. V. Hinkley, and O. E. Barndorff-Nielson, Eds, Likelihood, Time Series with Econometric and other Applications, pp. 1-67, Chapman Hall, London.

Taylor, S. J. (1986) Modelling Financial Time Series, Wiley, London.

Taylor, S. J. (2000) Consequences for Option Pricing of a Long Memory in Volatility, Working Paper, University of Lancaster.
24

Table 1: Autocorrelation Estimates for Squared and Absolute Returns at Different Power Transformations

| | k=0.25 | 0.5 | 0.75 | 1 | 1.25 | 1.5 | 1.75 | 2.0 |
|---|---|---|---|---|---|---|---|---|
| Squared | | | | | | | | |
| FTSE100 | | | | | | | | |
| > 0 | 4598 | 1010 | 340 | 242 | 195 | 171 | 153 | 25 |
| < 0 | 5 | 0 | 0 | 0 | 0 | 0 | 0 | 0 |
| UK Long Gilt | | | | | | | | |
| > 0 | 5722 | 2681 | 1360 | 2038 | 2208 | 1992 | 1468 | 840 |
| < 0 | 0 | 3 | 295 | 0 | 0 | 0 | 0 | 0 |
| Sterling | | | | | | | | |
| > 0 | 345 | 343 | 102 | 43 | 28 | 23 | 19 | 16 |
| < 0 | 3 | 5 | 0 | 0 | 0 | 0 | 0 | 0 |
| | | | | | | | | |
| Absolute | | | | | | | | |
| FTSE100 | | | | | | | | |
| > 0 | 3586 | 4598 | 3493 | 1010 | 434 | 340 | 281 | 242 |
| < 0 | 5 | 0 | 0 | 0 | 0 | 0 | 0 | 0 |
| UK Long Gilt | | | | | | | | |
| > 0 | 5722 | 5722 | 5720 | 2681 | 1255 | 1360 | 1721 | 2038 |
| < 0 | 0 | 0 | 0 | 3 | 326 | 295 | 75 | 0 |
| Sterling | | | | | | | | |
| > 0 | 273 | 345 | 394 | 343 | 194 | 102 | 63 | 43 |
| < 0 | 7 | 3 | 3 | 5 | 3 | 0 | 0 | 0 |

Notes: The critical values for each contract are ±0.008 (FTSE100), ±0.008 (UK Long Gilt) and ±0.012 (Sterling). The rows labelled > 0 and < 0 represent the number of autocorrelation values that are significantly greater than and less than zero respectively.

Table 2: APARCH Models for Five-minute UK Futures Returns

|  | FTSE100 | UK Long Gilt | Sterling |
|---|---|---|---|
| AR |  | 0.53 | 0.27 |
|  |  | (30.13)*** | (9.79)*** |
| MA |  | -0.57 | -0.43 |
|  |  | (-34.48)*** | (-16.77)*** |
| $\alpha_0$ | 0.01 | 7.10E-04 | 9.00E-07 |
|  | (3.13)*** | (8.35)*** | (1.53)* |
| $\alpha_1$ | 0.15 | 0.11 | 0.15 |
|  | (9.51)*** | (56.97)*** | (30.06)*** |
| $\beta_1$ | 0.82 | 0.86 | 0.62 |
|  | (39.15)*** | (310.81)*** | (73.38)*** |
| $\gamma_1$ | -0.09 | -0.08 | -0.09 |
|  | (-1.60) | (-4.21)*** | (-4.59)*** |
| $\delta$ | 1.07 | 0.47 | 1.17 |
|  | (9.14)*** | (-46.22)*** | (2.79)*** |
| Likelihood | 6162.87 | 384241 | 221382 |
| AIC | -2.05 | -768463 | -442746 |
| BIC | -2.04 | -768383 | -442672 |

Notes: Marginal significance levels using Bollerslev-Wooldridge standard errors are displayed by parentheses. A single asterisk denotes statistical significance at the 10%, two denotes statistical significance at the 5% level, while three denotes statistical significance at the 1% level. The t-statistics for the power coefficient, $\delta$, represents the value being significantly different from 1. Optimal models are chosen based on Akaike's (AIC) and Schwarz's (BIC) selection criteria.

Table 3: Autocorrelation Estimates for Rescaled Squared and Absolute Returns at Different Power Transformations

|  | k=0.25 | 0.5 | 0.75 | 1 | 1.25 | 1.5 | 1.75 | 2.0 |
|---|---|---|---|---|---|---|---|---|
| Squared |  |  |  |  |  |  |  |  |
| FTSE100 |  |  |  |  |  |  |  |  |
| > 0 | 570 | 473 | 301 | 205 | 154 | 122 | 99 | 79 |
| < 0 | 113 | 5 | 0 | 0 | 0 | 0 | 0 | 0 |
| UK Long Gilt |  |  |  |  |  |  |  |  |
| > 0 | 2224 | 956 | 1201 | 879 | 584 | 388 | 263 | 174 |
| < 0 | 8 | 109 | 0 | 0 | 0 | 0 | 0 | 0 |
| Sterling |  |  |  |  |  |  |  |  |
| > 0 | 241 | 96 | 63 | 24 | 8 | 4 | 2 | 2 |
| < 0 | 17 | 18 | 5 | 0 | 0 | 0 | 0 | 0 |
| Absolute |  |  |  |  |  |  |  |  |
| FTSE100 |  |  |  |  |  |  |  |  |
| > 0 | 4569 | 4704 | 3511 | 1009 | 434 | 340 | 281 | 242 |
| < 0 | 5 | 0 | 0 | 0 | 0 | 0 | 0 | 0 |
| UK Long Gilt |  |  |  |  |  |  |  |  |
| > 0 | 5722 | 5722 | 5720 | 2648 | 1254 | 1352 | 1692 | 1977 |
| < 0 | 0 | 0 | 0 | 3 | 315 | 292 | 75 | 0 |
| Sterling |  |  |  |  |  |  |  |  |
| > 0 | 1856 | 1643 | 1432 | 1123 | 665 | 266 | 129 | 48 |
| < 0 | 10 | 10 | 17 | 24 | 19 | 3 | 0 | 0 |

Notes: The critical values for each contract are ±0.008 (FTSE100), ±0.008 (UK Long Gilt) and ±0.012 (Sterling). The rows labelled > 0 and < 0 represent the number of autocorrelation values that are significantly greater than and less than zero respectively.

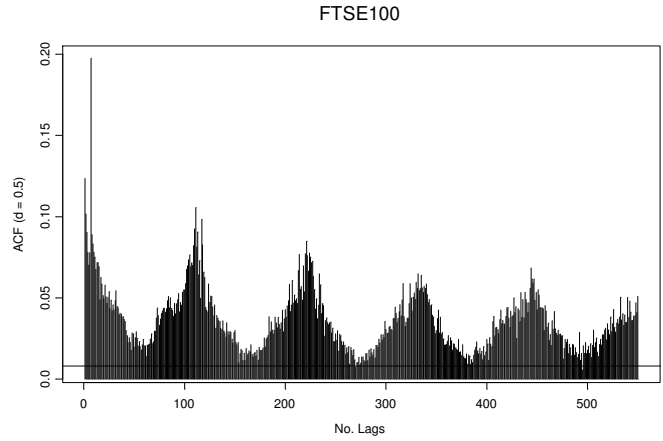

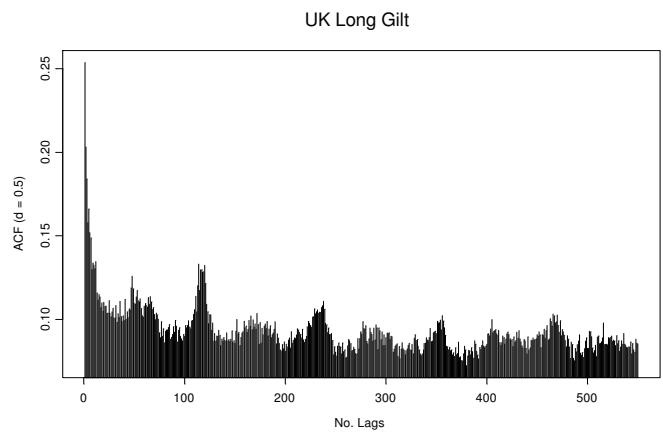

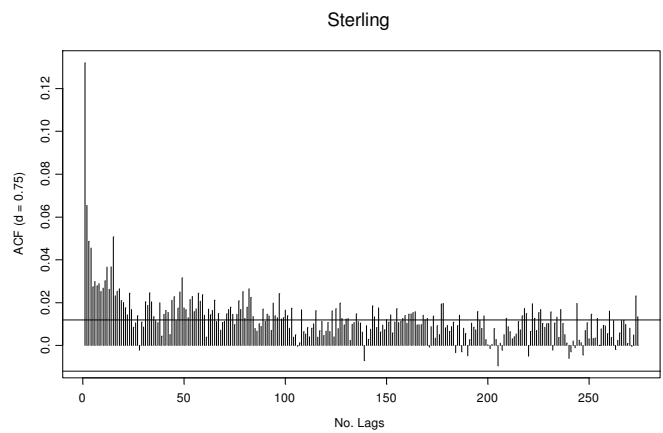

Figure 1: Plots of Autocorrelation Values across a trading week Absolute Returns showing the Long Memory Property at its Strongest. Confidence intervals are imposed on each plot.